\documentstyle[12pt,fleqn]{article}

\addtocounter{section}{-1}
\def\beeq{\begin{equation}}
\def\eneq{\end{equation}}
\def\beeqa{\begin{eqnarray}}
\def\eneqa{\end{eqnarray}}
\def\soc{{\rm C}_{60}}

\begin{document}
\vspace{0.8cm}
\begin{flushleft}
{\bf EXCITON AND LATTICE-FLUCTUATION EFFECTS\\
IN OPTICAL SPECTRA OF C$_{\bf 60}$}\\
\vspace{1.3cm}
KIKUO HARIGAYA AND SHUJI ABE\\
{\sl Electrotechnical Laboratory, Tsukuba, Ibaraki 305, Japan}
\end{flushleft}
\vspace{1.0cm}
\noindent
\underline{Abstract}
A theory of optical excitations by using a tight binding model with
long-range Coulomb interactions is developed.  The model is applied
to a $\soc$ molecule and a cluster, and is treated by the Hartree-Fock
approximation followed by a configuration interaction method.  Lattice
fluctuations are taken into account by a bond disorder model.
We first examine what strength of Coulomb interaction is appropriate
to describe the electronic structures observed by photo-electron
and optical absorption spectroscopy.  Then, we demonstrate that
the photo-excited states are mainly intramolecular (i.e. Frenkel)
excitons with small charge-transfer components.  We examine to
what extent the dipole forbidden transitions of a single $\soc$
molecule become dipole-allowed by lattice fluctuations or
intermolecular interactions.

\vspace{1.0cm}
\begin{flushleft}
\underline{INTRODUCTION}
\end{flushleft}

Optical experiments on fullerenes C$_N$ have revealed many interesting
properties associated with $\pi$ electrons delocalized on molecule surfaces.
The optical absorption spectrum of $\soc$,$^1$ provides basic information
about its electronic structure.   A large third-order
nonlinear susceptibility $\chi^{(3)}$ of the order $10^{-11}$esu
has been observed in the third-harmonic generation (THG) from $\soc$.$^{2,3}$

In order to analyze the optical properties and to clarify mechanisms
of the large nonlinearity, we have studied the linear absorption and
the THG of $\soc$ by using a tight binding model$^4$ and a model
with a long-range Coulomb interaction.$^5$  A free electron model
yields THG magnitudes which roughly coincide with the
experimental values of $\soc$,$^1$ while the calculated linear absorption
spectrum is not in satisfactory agreement with experiments.$^4$
If Coulomb interactions are taken into account, the absorption spectra
are in overall agreement with the experiment, although the magnitude
of $\chi^{(3)}$ becomes smaller by one order of magnitude.$^5$
We suggested that the observed values of $\chi^{(3)}$ may be explained
by taking local field corrections into account.

In this paper we first address the fundamental
question of what strength of Coulomb
interaction is appropriate in order to describe the electronic
structures observed by photo-electron and optical absorption spectroscopy.
The parameter of the bond disorder simulating the lattice fluctuations
is also specified.
Then, we study the character of the photo-excited states: whether
excitons are confined in a single molecule or they are distributed
among several molecules.  We also examine how dipole forbidden transitions
of the $\soc$ molecule
become dipole-allowed by lattice fluctuations and by intermolecular
interactions.

\begin{flushleft}
\underline{MODEL AND METHOD}
\end{flushleft}

In the calculation of an isolated $\soc$ molecule, the hopping
integrals, $t + (2/3) t'$ and $t - (1/3) t'$ are assigned for the
double and single bonds for $\pi$ electrons, respectively.  The
bond alternation $t' = 0.1t$ of a typical value is used in this
paper.  The average hopping integral $t$ is treated as an adjustable
parameter in the present study, being around 2eV.  In order to
simulate lattice fluctuation effects in the
molecule, Gaussian bond disorder is introduced in the hopping
integral.  The disorder strength is similar to that of the zero
point fluctuations.$^6$  The Pariser-Parr-Pople model with long range
Coulomb interactions is adopted
to describe exciton effects.  We use the Ohno potential
$W(r) = 1/\sqrt{(1/U)^2 + (r/r_0 V)^2}$ for the electron-electron
interactions, where $U$ is the strength of
the onsite interaction, $V$ means the strength of the long range
Coulomb interaction, and $r_0$ is the average bond length.

To discuss solid state effects, we consider a cluster of four molecules, which
corresponds to a unit cell of the simple cubic lattice in the low temperature
phase of the $\soc$ solid.  The periodic boundary condition is imposed.
In the system, one of pentagons on one $\soc$ molecule faces to
one of double bonds on a neighboring $\soc$ molecule.  We assume
weak intermolecular hopping integrals $t_{\rm w}$ between each of
the two double-bonded sites and its three closest sites on the facing
pentagon.  The distances of these intermolecular connections are fairly
large, about 3\AA, and $t_{\rm w}$ is estimated to be around
0.1 - 0.2eV.  We use $t_{\rm w} = 0.1t$ in the present paper.
In addition to the intermolecular hopping, we take into account
intermolecular electron-electron interactions by assuming
the Ohno potential also between sites on different molecules.
A proper long-distance cutoff is introduced for the potential to harmonize
with the cyclic boundary condition.

The model is treated by the Hartree-Fock approximation for the ground state
and the single-excitation configuration interaction (single CI)
method for excited states.  All the quantities with the energy dimension
are shown in the units of $t$.  We varied the parameters of Coulomb
interaction within $0 \leq V \leq U \leq 5t$, and we report here the
representative cases: $U = 4t$ and $V = 2t$, results of which
turn out to be in overall agreement with experiments.
In the calculation of the single molecule with bond disorder,
the average over 100 samples is taken.  This is sufficient for
obtaining smooth optical spectra.

\begin{flushleft}
\underline{OPTICAL ABSORPTION IN C$_{60}$ MOLECULES}
\end{flushleft}

Fig. 1(a) shows the calculated absorption spectrum, and Fig. 1(b)
shows the experimental data obtained by Ren et al.
for $\soc$ in a solution and a $\soc$ thin film$^1$ for comparison
assuming $t = 1.8$eV.  There are three main features at about $~2.0t$,
$~2.6 t$, and $~ 3.1t$ in the calculated spectrum.  Their relative positions
in energy agree well with the experimental data.
The relative oscillator strengths also agree well.  Furthermore,
the widths of the observed absorption peaks are well simulated by
the broadening in the bond disorder model.  In reality, the broad
absorption bands might contain fine structures due to many phonon
modes including intermolecular librations
and intramolecular vibrations.  However, they are not resolved in
the experiments presumably because they are smeared out by
some other weak broadening mechanisms (such as life-time broadening).
The disorder strength adopted here simulates well the overall
width due to the gross  contributions from the many phonon modes.

In the solid, the broadening is larger and a broad hump appears
around the energy 2.8eV $(=1.5t)$, as is shown in Fig. 1(b).
There is a tiny structure at 3eV in the corresponding region of
the absorption spectrum of $\soc$ in a solution (Fig. 1(b)).
We assume that the forbidden transitions in that energy region become
partially allowed due to lattice fluctuations or intermolecular
interactions.  If the lattice fluctuations are effective,
the effect can be simulated by the bond disorder model.
In Fig. 1(a), the absorption in the low energy part
multiplied by the factor 10 is also shown.  The several forbidden
transitions around $1.2t$ - $1.6t$ become allowed by the disorder,
giving rise to an absorption tail in the lower energy region.
The relative magnitude is similar to the solution data
shown in Fig. 1(b).  This might be also the origin of the 2.8eV
hump in the solid.  However, the strength of the absorption relative to the
main peaks in our theory is about one order of magnitude smaller than
that in the experiment.

\begin{flushleft}
\underline{OPTICAL ABSORPTION IN C$_{60}$ SOLIDS}
\end{flushleft}

We turn to calculations for $\soc$ solids.  Fig. 2(a) shows the density
of one-electron states calculated
in the Hartree-Fock approximation.  The occupied (unoccupied)
states are represented by the black (white) bars, respectively.
The distribution of states becomes broad due to the intermolecular
hopping, and the widths of the distribution are consistent with
the band calculations.$^7$  Thus, the hopping interaction $t_{\rm w} = 0.1t$
seems reasonable. The overall density of states agrees with the
photoemission and the inverse photoemission data$^8$
if we choose $t=1.8$eV.

Now, we discuss the optical spectra calculated by the single CI
method.  Fig. 2(b) shows the density of states for dipole allowed excitations
as a function of the excitation energy.  The hatched region indicates the
states for which the probability that the electron and the hole are
located in different molecules is larger than 0.5.  In other
words, these states consist mainly of charge transfer components.
Other states (nonhatched) correspond to primarily intramolecular
excitations, namely  Frenkel excitons.  Fig. 2(c) shows the
absorption spectrum where the Lorentzian broadening $\Gamma = 0.03 t$
is used.  The absorption in the low energy region $1.6t$ - $1.9 t$ is
slightly enhanced from those in Fig. 1(a) owing to the intermolecular
interactions, although the gross features of the spectrum remain unchanged.
This is consistent with the experimental data shown in Fig. 1(b).
We would like to stress that the states which give rise to
relatively large oscillator strengths are the states with small charge
transfer components.  The states with large charge transfer
components give very small oscillator strengths.  Thus,
we conclude that the photoexcited states are mainly
Frenkel-exciton-like states.  This is also in accord with the fact
that the excitation energies of such states are largely shifted to
low energies from the one-electron excitation energies obtained
at the Hartree-Fock level.  For example, the absorption peak at
about $1.7t$ in Fig. 2(b) originates from the one-electron
transitions between the highest occupied valence band at about
$-1.2t$ and the second lowest unoccupied band at about $1.1t$,
the distance between them being about $2.2t$.

The states near the energy $1.5t$ in Fig. 2(b), which are forbidden
in the single molecule, become dipole allowed owing to the intermolecular
interactions.  However, these states give very small contributions
to the spectrum.  Therefore, it seems again difficult to explain the
magnitude of the observed absorption below 3eV in the solid by means of
the simple intermolecular interactions considered here.

\begin{flushleft}
\underline{SUMMARY}
\end{flushleft}

We have developed a theory of optical excitations of $\soc$
by using a tight binding model with long-range Coulomb interactions.
The parameters, $U \sim 4t$ and $V \sim 2t$, in the Ohno potential
turned to be appropriate to describe the absorption spectra.
The strength of the bond disorder $t_{\rm s} \sim 0.1t$ seems
reasonable to simulate lattice fluctuation effects.  Next, we have
demonstrated that the photo-excited states are mainly intramolecular
(i.e. Frenkel) excitons with small charge-transfer components.
We have also examined how dipole-forbidden transitions of a single
$\soc$ molecule at low energies become dipole-allowed by solid state
effects or lattice fluctuations.  Both effects give rise to a theoretical
oscillator strength of about one order of magnitude smaller than in
experiments  for the solid.  Combination of
the two mechanisms or inclusion of other origins, for example,
orientational disorder, may be necessary.

\begin{flushleft}
\underline{REFERENCES}
\end{flushleft}

\noindent
1. S. L. Ren, Y. Wang, A. M. Rao, E. McRae, J. M. Holden, T. Hager,
KaiAn Wang, W. T. Lee, H. F. Ni, J. Selegue, and P. C. Eklund,
\underline{Appl. Phys. Lett.}, \underline{59}, 2678 (1991).\\
2. J. S. Meth, H. Vanherzeele, and Y. Wang, \underline{Chem. Phys. Lett.},
\underline{197}, 26 (1992).\\
3. Z. H. Kafafi, J. R. Lindle, R. G. S. Pong, F. J. Bartoli,
L. J. Lingg, and J. Milliken, \underline{Chem. Phys. Lett.},
\underline{188}, 492 (1992).\\
4. K. Harigaya and S. Abe, \underline{Jpn. J. Appl. Phys.},
\underline{31}, L887 (1992).\\
5. K. Harigaya and S. Abe, \underline{J. Lumin.} (to be published).\\
6. K. Harigaya, \underline{Phys. Rev. B}, \underline{48}, 2765 (1993).\\
7. S. Saito and A. Oshiyama, \underline{Phys. Rev. Lett.},
\underline{66}, 2637 (1991).\\
8. J. H. Weaver, \underline{J. Phys. Chem. Solids}, \underline{53},
1433 (1992).\\

\mbox{}

\noindent
FIGURE 1  Optical absorption spectra for a $\soc$ molecule shown in
arbitrary units.  In (a), the spectrum is calculated with the parameters
$U = 4t$, $V = 2t$, and the bond disorder of the strength $t_s = 0.09t$.
The abscissa is scaled by $t$.  (b) The experimental
spectra (Ref. 1) of molecules in a solution (solid line)
and of a $\soc$ film (dotted line).  We use $t = 1.8$eV.

\mbox{}

\noindent
FIGURE 2 (a) Density of states of the cluster by the Hartree-Fock
approximation.  The occupied (unoccupied) states are shown by
the black (white) bars.  (b) Density of states for dipole allowed
excitations as a function of the excitation energy.  (c) The
absorption spectrum where the Lorentzian broadening $\Gamma = 0.03 t$
is used.

\end{document}